\documentclass[
 aip,
 amsmath,amssymb,
 reprint,%
]{revtex4-1}

\usepackage{graphicx}
\usepackage{dcolumn}
\usepackage{bm}
\usepackage{physics}
\usepackage{xcolor}

\usepackage[utf8]{inputenc}
\usepackage[T1]{fontenc}
\usepackage{etoolbox}

\usepackage{color}

\makeatletter
\def\@email#1#2{%
 \endgroup
 \patchcmd{\titleblock@produce}
  {\frontmatter@RRAPformat}
  {\frontmatter@RRAPformat{\produce@RRAP{*#1\href{mailto:#2}{#2}}}\frontmatter@RRAPformat}
  {}{}
}%
\makeatother
\begin{document}

\title{Predicting the Electronic Density Response of Condensed-Phase Systems to Electric Field Perturbations}
\author{Alan M.~Lewis}
\email{alan.lewis@mpsd.mpg.de}
\affiliation{Max Planck Institute for the Structure and Dynamics of Matter, Luruper Chaussee 149, 22761 Hamburg, Germany}
\author{Paolo Lazzaroni}
\affiliation{Max Planck Institute for the Structure and Dynamics of Matter, Luruper Chaussee 149, 22761 Hamburg, Germany}
\author{Mariana Rossi}
\email{mariana.rossi@mpsd.mpg.de}
\affiliation{Max Planck Institute for the Structure and Dynamics of Matter, Luruper Chaussee 149, 22761 Hamburg, Germany}

\date{\today}

\begin{abstract}
We present a local and transferable machine learning approach capable of predicting the real-space density response of both molecules and periodic systems to external homogeneous electric fields. 
The new method, SALTER, builds on the Symmetry-Adapted Gaussian Process Regression SALTED framework. SALTER requires only a small, but necessary, modification to the descriptors used to represent the atomic environments. We present the performance of the method on isolated water molecules, bulk water and a naphthalene crystal. Root mean square errors of the predicted density response lie at or below 10\% with barely more than 100 training structures. Derived quantities, such as polarizability tensors and even Raman spectra further derived from these tensors show a good agreement with those calculated directly from quantum mechanical methods. Therefore, SALTER shows excellent performance when predicting derived quantities, while retaining all of the information contained in the full electronic response. This method is thus capable of learning vector fields in a chemical context and serves as a landmark for further developments.
\end{abstract}

\maketitle


Machine-learning (ML) tools are now widely used in computational chemistry to reduce the cost of simulations and aid the interpretation of data.\cite{Butler2018,Ceriotti2019,Carleo2019,Deringer2021} ML models are often employed to predict energies and forces,\cite{Thompson2015,Kamath2018,Drautz2019,Deringer2019,Behler2021} which are then used to drive molecular dynamics simulations.\cite{Noe2020,Friederich2021,Unke2021,Musaelian2023} This has opened the door to \emph{ab initio} quality simulations on timescales and system sizes which are completely inaccessible by conventional AIMD approaches.\cite{Sosso2012,Botu2017,Maresca2018,Deringer2018,Musaelian2023} However, this approach comes at the cost of losing access to other properties directly related to the electronic density.

These must be obtained by some independent method, which triggered the development of many recent models.\cite{Darley2008,Unke2019,Raimbault2019,Veit2020,Pinheiro2020,Sun2022} We have recently developed SALTED, a machine learning method which directly predicts the real-space electronic density of molecular and condensed phase systems.\cite{Lewis2021,Grisafi2022} SALTED may be applied in tandem with machine-learned potentials, providing access to every ground-state electronic-structure property which would be available in a traditional AIMD calculation.

In this Communication, we present SALTER (Symmetry-Adapted Learning of Three-dimensional Electron Responses). SALTER builds on SALTED to predict the static real-space \emph{response} of the electron density of a molecule or material to a homogeneous electric field. This response defines the global dielectric susceptibility of the system, which is needed to simulate many flavors of vibrational Raman and sum-frequency spectra,\cite{Hendra1969,McQuarrie1976,Morita2000,Morita2002,Auer2008,Lee2019, Raimbault2019a} and defines the high-frequency electronic screening which impacts a wide range of static and transport properties of molecules and materials.~\cite{Sohier2015, Cudazzo2011, Skone2014, Brawand2016,Shang2018}   

The computational burden of calculating this electronic response is considerable -- the most commonly employed method, density-functional perturbation theory (DFPT),\cite{Baroni2001} scales similarly to the underlying DFT calculation with respect to system size, but with a 4-10 times larger prefactor.\cite{Shang2018,Andrade2007} As such, this calculation is a bottleneck in assessing the dielectric properties of systems along long molecular dynamics trajectories and large system sizes.

As in our previous work,\cite{Lewis2021,Grisafi2022} SALTER is based on an atom-centred expansion of the density response, with the coefficients of the expansion predicted using the symmetry adapted Gaussian process regression (SA-GPR) approach.\cite{Grisafi2018} This produces a method which is local, transferable, and capable of treating molecular and periodic systems on the same footing.\cite{Lewis2021} The key modification introduced in this work is the inclusion of a ``dummy'' atom in the representation of each atomic environment, which encodes information about the direction of the applied perturbation in a simple way, while retaining all of the symmetry properties of the representation. This modification is significantly simpler than formally including the applied field in the descriptor, which would require taking the tensor product of each symmetry-adapted kernel with a rank-1 tensor describing the applied field.

 We start by briefly presenting the method. We follow the same general approach outlined in our previous work learning the electron density, $\rho$,\cite{Lewis2021,Grisafi2022} and highlight the key differences to the previous method. For a more detailed account we refer the reader to Refs. \citenum{Lewis2021} and \citenum{Grisafi2022}. The density response $\rho^{(1)}_{\beta}$ to a field applied along a Cartesian axis $\beta$ can be expanded using a set of auxiliary basis functions $\phi$, such that
\begin{equation}
\begin{aligned}
& \frac{d\rho(\mathbf{r})}{de_\beta} \equiv \rho^{(1)}_\beta(\mathbf{r})  \\ & \approx \tilde{\rho}^{(1)}_\beta (\mathbf{r}) 
= \sum_{i,\sigma,\mathbf{U}} c_{i\sigma\beta} \phi_{i\sigma}(\mathbf{r} - \mathbf{R}_i + \mathbf{T}(\mathbf{U})). 
\end{aligned}
\label{eq:approx_rho1}
\end{equation}
Here $\mathbf{R}_i$ is the position of atom $i$, the basis function $\phi_{i\sigma}$ is centered on atom $i$ and may be written as the product of a radial part $R_n(r)$ and a real spherical harmonic $Y_{\lambda\mu}(\theta,\phi)$; for brevity we use a composite index $\sigma \equiv (an\lambda\mu)$, where $a$ labels the atomic species. $\mathbf{T}(\mathbf{U})$ is a translation vector to a point removed from the central reference unit cell by an integer multiple $\mathbf{U} = (U_x,\, U_y,\, U_z)$ of the lattice vectors, present only when the system under consideration is periodic. The optimal coefficients for this expansion can be found by minimising the loss function 
\begin{equation}
\epsilon(\mathbf{c}_\beta^{\text{RI}}) = \int_{u.c.} d\mathbf{r} \left| \tilde{\rho}^{(1)}_\beta(\mathbf{r};\mathbf{c}_\beta^{\text{RI}}) - \rho^{(1)}_\beta(\mathbf{r}) \right|^2,
\label{eq:loss_func1}
\end{equation}
which yields
\begin{equation}
\mathbf{c}^{\text{RI}}_\beta = \mathbf{S}^{-1}\mathbf{w}_\beta^{(1)}.
\label{eq:invert}
\end{equation}
Here $\mathbf{S}$ is the overlap matrix of the periodic or non-periodic basis functions, and $\mathbf{w}_\beta^{(1)}$ is a vector of the projections of the self-consistent density response $\rho^{(1),\text{QM}}_\beta(\mathbf{r})$ onto the basis,
\begin{equation}
    w^{(1)}_{i\sigma\beta} = \sum^{\mathbf{U}_\text{cut}}_\mathbf{U}\braket{\phi_{i\sigma}(\mathbf{r} - \mathbf{R}_i + \mathbf{T}(\mathbf{U}))}{\rho^{(1)}_\beta(\mathbf{r})}_\text{u.c.}
\label{eq:w}
\end{equation}
These optimal coefficients $\mathbf{c}^{\text{RI}}_\beta$ serve as the references against which the accuracy of coefficients predicted by the ML model $\mathbf{c}_\beta^{\text{ML}}$ are evaluated.

In order to predict $\mathbf{c}^{\textrm{ML}}_\beta$, we approximate them as a linear combination of the regression weights $b_{j\sigma\beta}$ associated with $M$ reference environments, and a covariant kernel $\operatorname{k}_{\sigma\beta}(A_i,M_j)$ which describes the similarity between the reference environments $\left\{M_j\right\}$ and the atomic environments which comprise the target structure $\left\{A_i\right\}$:
\begin{equation}
    c^{\textrm{ML}}_{i\sigma\beta} \approx \sum_j^{M} b_{j\sigma\beta} \operatorname{k}_{\sigma\beta}(A_i,M_j) .
\label{eq:coeffs_approx}
\end{equation}
Note that here $c^{\textrm{ML}}_{i\sigma\beta}$ and $b_{j\sigma\beta}$ are vectors of length $2\lambda + 1$, and $\operatorname{k}_{\sigma,\beta}(A_i,M_j)$ is a square matrix of the same dimension. Defining and minimizing a loss function analogous to Eq.~\eqref{eq:loss_func1} allows us to determine the regression weights $\mathbf{b}_{\beta}$:
\begin{equation}
\begin{split}
\epsilon(\mathbf{b}_\beta) = & \sum_{A=1}^N \int_{u.c.} d\mathbf{r} \left| \tilde{\rho}^{(1),\text{ML}}_\beta(\mathbf{r};\mathbf{b}_\beta) - \rho^{(1),\text{RI}}_\beta(\mathbf{r}) \right|^2 \\
& + \eta\,\mathbf{b}^T\mathbf{K}_{MM}\mathbf{b}\, .
\end{split}
\label{eq:loss_func2}
\end{equation}
Here the index $A$ runs over each of the $N$ structures in the training set, the matrix $\mathbf{K}_{MM}$ couples the set of reference atomic environments to one another, and we introduce the hyperparamter $\eta$ to control the regularization of the minimization. In all the examples presented here, we use a regularization parameter of $\eta = 10^{-8}$, finding only a weak dependence of the observed error on $\eta$. Eq.~\eqref{eq:loss_func2} is solved iteratively using the conjugate gradient algorithm; for a detailed description of the implementation the reader is referred to Ref.~\citenum{Grisafi2022}.

The key difference between SALTER and SALTED lies in the similarity kernel $\operatorname{k}_{\sigma\beta}(A_i,M_j)$. In Refs.~\citenum{Lewis2021} and \citenum{Grisafi2022}, kernels are constructed from the $\lambda$-SOAP representation of the atomic environments $A_i$ and $M_j$.\cite{Grisafi2018} This captures the symmetry covariant transformations of the spherical harmonics under rotation; the coefficients must also possess this symmetry since they expand auxiliary basis functions whose angular parts are spherical harmonics. As such, they are independent of the laboratory frame. However, in the present case the similarity kernel $\operatorname{k}_{\sigma\beta}(A_i,M_j)$ must retain all of the properties exploited in previous work, while also including the laboratory frame direction along which the perturbing field is applied.

\begin{figure}[t]
    \centering
    \includegraphics[width=0.45\textwidth]{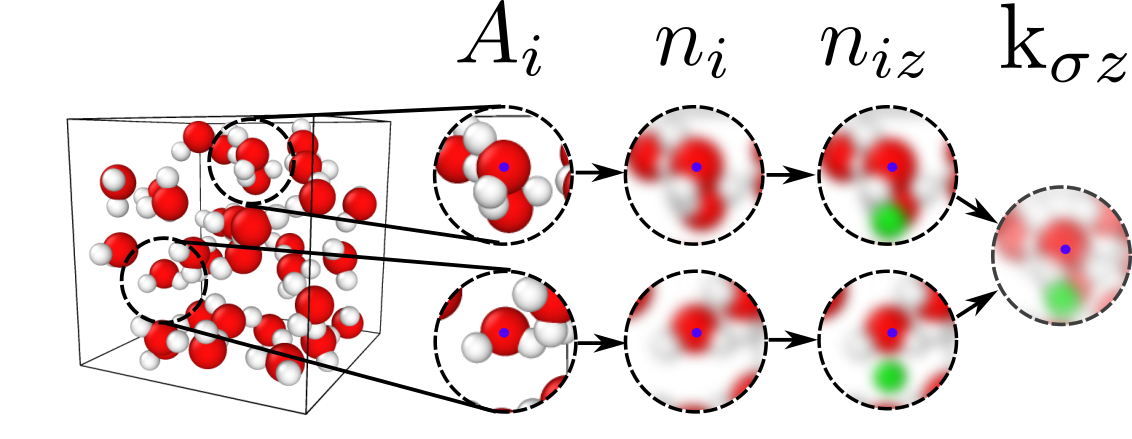}
    \caption{ An illustration of the modified $\lambda$-SOAP descriptors and kernels. Environments $A_i$ are defined by a cutoff radius around a central atom $i$, labelled with a blue dot. The usual SOAP environments $n_i$ (see Eq.~\eqref{eq:environment1}) are constructed from a Gaussian function centred on every atom within the cutoff radius. Our modified environments $n_{i\beta}$ (see Eq.~\eqref{eq:environment2}) add a Gaussian corresponding to an additional ``dummy'' atom, shown here in green, which encodes the direction of the applied field in the descriptor. The $\lambda$-SOAP kernels are then built from the symmetry-adapted overlap of these modified environments, as in Eq.~\eqref{eq:SOAP}. }
    \label{fig:dummy}
\end{figure}

Our solution is simple. An additional ``dummy'' atom of atomic number zero is added to every atomic environment on the $\beta$ axis, differentiating this axis from the others. The kernels are then calculated using $\lambda$-SOAP representations of these modified atomic environments in the normal way. This dummy atom is then naturally included in the kernels for all values of $\lambda$, ensuring that the kernels retain all of the required symmetry properties, while including (symmetry-adapted) information about the laboratory frame direction of the applied field. This idea is illustrated in Fig.~\ref{fig:dummy}, and  defined mathematically below.

The unmodified $\lambda$-SOAP kernel is defined as\cite{Grisafi2018}
\begin{equation}
    \operatorname{k}_{\sigma}(A_i,M_j) = \int \textrm{d} \hat{R} \textbf{D}^\lambda(\hat{R}) \left| \int n_i(\textbf{r}) n_j (\hat{R}\textbf{r}) \textrm{d} \textbf{r} \right|^2 ,
\label{eq:SOAP}
\end{equation}
where the first integral is performed over all rotations of order $\lambda$, and $\textbf{D}^\lambda(\hat{R})$ is the Wigner matrix corresponding to rotation $\hat{R}$. The atomic environment around atom $i$ is described as a sum of Gaussian functions of width $\nu$ centred at the position of every atom $\textbf{R}_j$ which lies within some radius $r_c$ of the central atom:
\begin{equation}
    n_i(\textbf{r}) = \sum_j g_\nu(\textbf{r} - \mathbf{R}_j) \, \Theta (r_c - |\textbf{R}_j - \mathbf{R}_i|) .
\label{eq:environment1}
\end{equation}
$\Theta(x)$ is the Heaviside step function. To include information about the direction of an applied field, we include a ``dummy'' atom in the atomic environment, adding an additional term to Eq.~\eqref{eq:environment1}:
\begin{equation}
    n_{i\beta}(\textbf{r}) = \sum_j g_\nu(\textbf{r} - \mathbf{R}_j) \Theta (r_c - |\textbf{R}_j - \mathbf{R}_i|) + g_\zeta(\textbf{r} - x \hat{\textbf{r}}_{\beta} r_c)
\label{eq:environment2}.
\end{equation}
Here $x$ is a parameter between 0 and 1 specifying the position of the dummy atom and $\hat{\textbf{r}}_{\beta}$ is a unit vector in the Cartesian direction $\beta$. Note that the width of the Gaussian associated with the dummy atom is not required to be the same as that associated with the physical atoms; in general $\zeta \neq \nu$. The $\lambda$-SOAP kernel $\operatorname{k}_{\sigma\beta}$ which accounts for the direction of the applied field is then obtained by replacing $n_i(\textbf{r})$ with $n_{i\beta}(\textbf{r})$ in Eq.~\eqref{eq:SOAP}.

 Throughout this work, we define the \% RMSE across the  datasets as
\begin{equation}
    \frac{\textrm{\% RMSE}}{100} = \sqrt{ \frac{ \frac{1}{N} \sum_A^{N} \int_{u.c.} d\mathbf{r} \left| \tilde{\rho}_{A\beta}^{(1),\text{ML}}(\mathbf{r}) - \tilde{\rho}_{A\beta}^{(1),\text{RI}}(\mathbf{r}) \right|^2 \\}{\frac{1}{N-1} \sum_A^{N} \Delta \bar{\mathbf{c}}_{A\beta}^T \Delta \bar{\mathbf{w}}^{(1)}_{A\beta}}}.
\end{equation}
Here the sums run over all of the structures $A$ in the training set, and the denominator is the standard deviation of the density response across the training set, with $\Delta \bar{\mathbf{c}}_A = \mathbf{c}_A - \bar{\mathbf{c}}$ and $\Delta \bar{\mathbf{w}}^{(1)}_A = \mathbf{w}^{(1)}_A - \bar{\mathbf{w}}^{(1)}$; $\bar{\mathbf{c}}$ and $\bar{\mathbf{w}}^{(1)}$ are the mean values for the vectors of coefficients and density projections in the data set.

In all of the following we used the FHI-aims software package\cite{Blum2009} to obtain the training data $\mathbf{S}$ and $\mathbf{w}_\beta^{(1)}$ and reference density responses $\rho^{(1),\textrm{QM}}_\beta$, using density functional theory\cite{Hohenberg1964,Kohn1965} employing the LDA functional.\cite{Hohenberg1964} In all cases ``light'' basis sets were used, and the auxiliary basis functions introduced in Eq.~\eqref{eq:approx_rho1} are constructed by taking all possible products of these basis functions and then eliminating linear dependencies from the resulting basis set.\cite{Ren2012} The molecular dynamics simulations were performed using the i-PI code.\cite{Kapil2019} We note that the ``light'' settings used are not the most accurate for the prediction of these properties, but our purpose here is simply testing the quality of the ML model, which can be achieved at this level of accuracy. Increasing the model accuracy only requires calculating new training data, which, as we shall see below, is not very costly. The training and reference data are available from the repositories listed in the SM.

\begin{figure}
\includegraphics[width=0.2\textwidth]{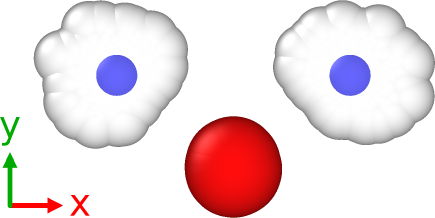}
\vspace{0.5cm}
\par
\begin{tabular}{c | c c c } 
 Kernel & $x$ & $y$ & $z$ \\
 \hline
 $\operatorname{k}_{\sigma}$ &  115.5 & 7.93 & 99.9 \\
 $\operatorname{k}_{\sigma\beta}$  & 1.04 & 0.63 & 0.96  \\
\end{tabular}
\caption{ Above: A visualisation of the 1000 water molecule dataset in which all structures are overlayed. Each molecule lies in the $xy$ plane, with the O atom at the origin; the positions of the H atoms are varied around the equilibrium geometry, which is highlighted in blue and whose centre of mass lies on the $y$ axis. Below: The \% RMSE across the testset of 750 structures, using $\lambda$-SOAP kernels built using atomic environments which exclude (first row) and include (second row) a dummy atom. }
\label{fig:isolated}
\end{figure}

First, we tackle the simple example of an isolated water molecule. We learn the density response to an electric field for a dataset of 1000 distorted configurations taken from Ref.~\citenum{Grisafi2018}, using both the usual kernels $\operatorname{k}_{\sigma}$ and the modified kernels $\operatorname{k}_{\sigma\beta}$. Each molecule in the dataset is aligned to a reference molecule, as shown in Fig.~\ref{fig:isolated}. We use the SOAP parameters $r_c = 4.0$ \AA, $\nu = 0.3$ \AA, as is commonly used to describe water,\cite{Grisafi2018} and chose $\zeta = \nu$ and $x = 0.9$ to define the dummy atom. The choice of these parameters will be discussed in more detail below, and makes little difference to the results in this simple example. The dataset of was split into a training set of 250 and a testset of 750 structures; $M=300$ reference environments were used.

The accuracy of SALTER density responses for this simple dataset are summarised in the table in Fig.~\ref{fig:isolated}. The first row shows that while aligning the molecules is sufficient to allow reasonably accurate learning of the density response to a field applied along the $y$ axis, the learning entirely fails for fields applied along the $x$ or $z$ axis. This can be explained by symmetry: since the molecules lie in the $xy$ plane, the  SOAP descriptors are invariant to reflections in that plane. As a result, there is no differentiation between a field applied in the $+z$ and $-z$ direction, resulting in errors approaching 100\%. By contrast, once the dummy atom is introduced, this ambiguity is removed; the direction along which the field is applied is clearly defined. As a result, we see similar errors regardless of the field direction, which are around 1\% in every case. Indeed, we also see an improvement in the learning of the response to a field in the $y$ direction when using the modified kernels, since the applied field direction is made more explicit than simply being inferred from the aligned geometries.

This simple example also displays a feature of SALTER. Since the molecules are aligned, the Cartesian axes are nonequivalent. As a result, a separate machine-learning model must be trained for each non-equivalent direction. We argue that this additional cost is offset by the conceptual simplicity of the approach, the computational simplicity of calculating the descriptors, and that this drawback only applies in the specific case of nonequivalent axes. We also stress that it is not necessary to align systems for this approach to succeed; the alignment here has a purely pedagogical purpose.


Having established that our approach works in principle, we turn to a more realistic example: a cubic water cell of side 9.67 \AA~containing 32 water molecules. Our dataset consists of 500 configurations, divided into a training set of 400 structures and a test set of 100 structures; this is the same dataset as used in Ref.~\citenum{Grisafi2022}. Because there are now physical atoms distributed throughout each spherical atomic environment, it is not clear \emph{a priori} where the dummy atom should be placed within the environment, or what the optimal values for $x$ and $\zeta$ in Eq.~\eqref{eq:environment2} would be. We performed a search of the parameter space using a subset of 50 structures from the full training set, finding the lowest errors when $x=0.3$ and $\zeta = 1.0$ \AA. Full details of this optimisation are provided in the Supplementary Material.

\begin{figure}[t]
    \centering
    \includegraphics[width=\columnwidth]{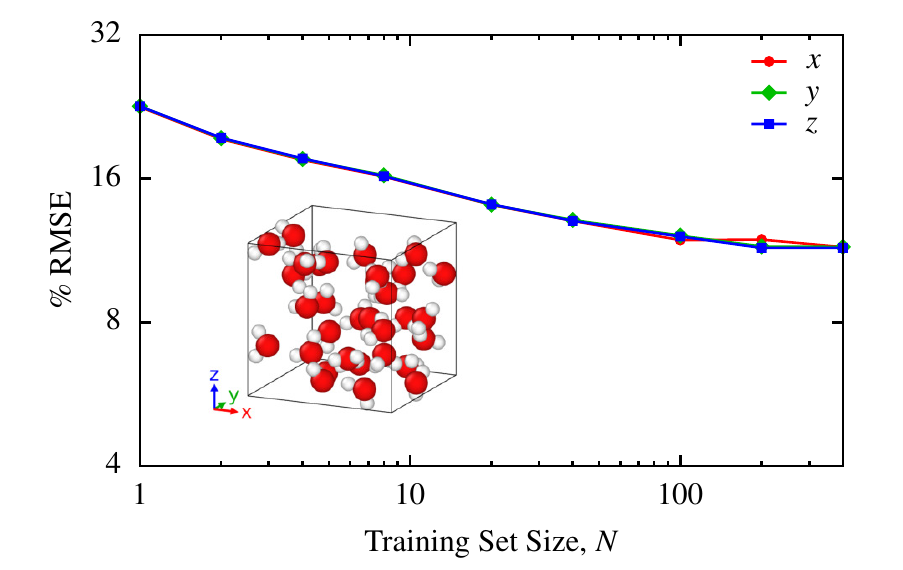}
    \caption{ Learning curves for the density response to a field along each Cartesian direction for the bulk water dataset, $M=3000$. }
    \label{fig:lcs_xyz}
\end{figure}

The learning curves for the density response to a field applied along each Cartesian axis are shown in Fig.~\ref{fig:lcs_xyz}. Learning improves monotonically with increasing training set size, with the error reaching a plateau of around 12\% for $N \ge 100$. We have used $M=3000$ reference environments for these calculations (see convergence with respect to $M$ in the SM). As the Cartesian directions are all equivalent, the machine learning model trained on the density response to a field applied along the $x$ axis is also used to predict the response to a field along the $y$ and $z$ axes; the errors observed are very similar regardless of the field direction, as would be expected by symmetry.

The learning curve for this dataset saturates at a relatively large \% RMSE: around three times larger than the error in the predicted densities we reported in our previous work.\cite{Grisafi2022} The reason for this could be a deficiency of the descriptor, or a lack of quality in the training data. Our descriptors are short ranged: any correlations in the response of the density to a field which are longer-ranged than the 4 \AA~cutoff radius of the atomic environment will be lost, limiting the accuracy of predictions. Alternatively, more accurate predictions may be possible if the training data was obtained using a larger basis set than the ``light'' basis sets found in AIMS, since this both allows a more accurate expansion of the density response, and reduces the error in the training data which in turn reduces the noise in the machine-learning model.

We then tested the performance of SALTER on a derived quantity from the electronic response, namely the dielectric susceptibility tensor. This quantity is related to the polarizability of individual molecules and its elements are  defined by~\cite{Shang2018}
\begin{equation}
   \alpha_{\gamma\delta} = - \int_{\textrm{u.c.}} r_{\gamma}  \rho^{(1)}_\delta(\bm{r}) \textrm{d} \bm{r} .
    \label{eq:alpha}
\end{equation}
Since we are considering a periodic system, when evaluating Eq.~\eqref{eq:alpha} we exploit the commutation relation between the unperturbed Hamiltonian and the position operator to avoid an ill-defined result.\cite{Baroni2001, Shang2018} We denote the tensor defined by the density response found by a self-consistent DFPT calculation as $\bm{\alpha}^{\textrm{QM}}$, while $\bm{\alpha}^{\textrm{ML}}$ is that derived from the predicted density response, and the error in $\bm{\alpha}^{\textrm{ML}}$ relative to $\bm{\alpha}^{\textrm{QM}}$ provides a measure of the accuracy of the predicted density responses. Table \ref{tab:bulk_pol} shows the errors in different elements $\bm{\alpha}^{\textrm{ML}}$ derived from the density response predicted by the machine learning model with $N = 400,\,M = 3000$. In all cases errors are around 5\%; this compares favourably to previous direct machine-learning predictions of $\bm{\alpha}^{\textrm{ML}}$, which typically find errors $>10\%$ for models trained on a few hundred structures, reducing to 5\% only when trained on a few thousand structures.\cite{Raimbault2019,Wilkins2019a,Shang2021} Therefore, for a given accuracy this indirect method of predicting $\bm{\alpha}^{\textrm{ML}}$ reduces the computational effort required to generate training data by an order of magnitude relative to a direct prediction of $\bm{\alpha}^{\textrm{ML}}$.

The smaller errors in $\bm{\alpha}^{\text{ML}}$ than in $\bm{\rho}^{(1), \text{ML}}$  can be understood from Eq.~\eqref{eq:alpha}. If errors in the density response accumulate in regions of space which do not contribute significantly to the integral, one would expect to see lower errors in $\bm{\alpha}^{\text{ML}}$ than the density response from which it is derived. Indeed, we discussed a similar dependence of the error in properties derived from electron density on the spatial distribution of errors in the electron density itself in Ref.~\citenum{Grisafi2022}.

\begin{table}
\centering
\setlength{\tabcolsep}{5pt}
\begin{tabular}{c | c c c c c c  } 
 Component & $xx$ & $yy$ & $zz$ & $xy$ & $xz$ & $yz$ \\
 \hline
 \% RMSE & 5.03 & 4.88 & 4.39 & 5.11 & 4.91 & 4.31 \\
\end{tabular}
\caption{\% RMSE in each component of the dielectric susceptibility derived from the density responses predicted by a machine learning model trained using $N = 400,\,M = 3000$ for the bulk water test set.}
\label{tab:bulk_pol}
\end{table}


Finally, we show the performance of SALTER on a molecular crystal and on a quantity which is ``twice removed'' from the target prediction of the model, namely the vibrational Raman spectrum. We take the P2$_1$/a naphthalene crystal at 80 K. We used five 10 ps NVE \emph{ab initio} molecular dynamics trajectories of a $2 \times 2 \times 1$ crystal supercell (8 molecules) using the PBE functional\cite{Perdew1996} with many-body dispersion corrections.\cite{Tkatchenko2012,Ambrosetti2014} To train the machine learning model, we selected 100 structures from each of these trajectories, dividing these structures into a training set of 400 configurations and a test set of the remaining 100. The learning curves for the density response to a field applied along each Cartesian direction are shown in Fig.~\ref{fig:naph_lcs_xyz}. These curves are converged with respect to $M$, as demonstrated in the Supplementary Material, and we see no significant difference in performance when predicting the response to a field applied along each Cartesian axis. We note that the dataset is relatively homogeneous: a single training structure is sufficient to produce a ML model accurate to within 12\%, and no improvement is seen when the training set size is increased beyond 100.

We then used the predicted density responses to calculate $\bm{\alpha}^{\text{ML}}$ for each structure in the test set. Curiously, as shown in Table \ref{tab:naph_pol}, the errors are significantly larger than those found in bulk water (Table \ref{tab:bulk_pol}), despite the \% RMSE of the density response itself being significantly lower for naphthalene. Furthermore, the errors in the components of $\bm{\alpha}^{\text{ML}}$ derived from a response to a field applied in the $z$ direction are the largest, despite the errors in this response being the lowest (see crystal cell orientation in the inset of Fig.~\ref{fig:naph_lcs_xyz}). This counter-intuitive result is again due to the fact that the errors in $\bm{\alpha}^{\text{ML}}$ depend not simply on the magnitude of the error but also its spatial distribution. Specifically, for a given magnitude of error, the greater the distance between the regions of positive and negative error, the greater the error in the dielectric susceptibility derived from the density response -- and this is what we observe along the $z$ direction in this crystal. This is illustrated in more detail in the SM.

\begin{figure}[t]
    \centering
    \includegraphics[width=\columnwidth]{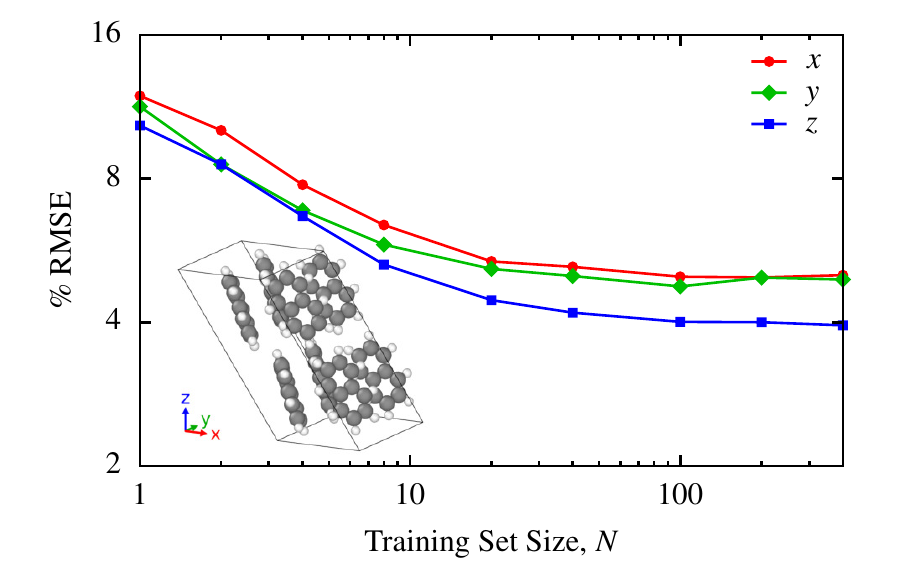}
    \caption{ Learning curves for the density response to a field along each Cartesian direction for the naphthalene test set, $M=2000$. }
    \label{fig:naph_lcs_xyz}
\end{figure}

\begin{table}
\centering
\setlength{\tabcolsep}{5pt}
\begin{tabular}{c | c c c c c c  } 
 Component & $xx$ & $yy$ & $zz$ & $xy$ & $xz$ & $yz$ \\
 \hline
 \% RMSE & 9.13 & 14.40 & 19.74 & 9.38 & 12.81 & 20.11 \\
\end{tabular}
\caption{ \% RMSE in each component of the dielectric susceptibility derived from the density responses predicted by machine learning models trained using $N = 400,\,M = 3000$ for the naphthalene test set. }
\label{tab:naph_pol}
\end{table}

\begin{figure}[t]
    \centering
    \includegraphics[width=\columnwidth]{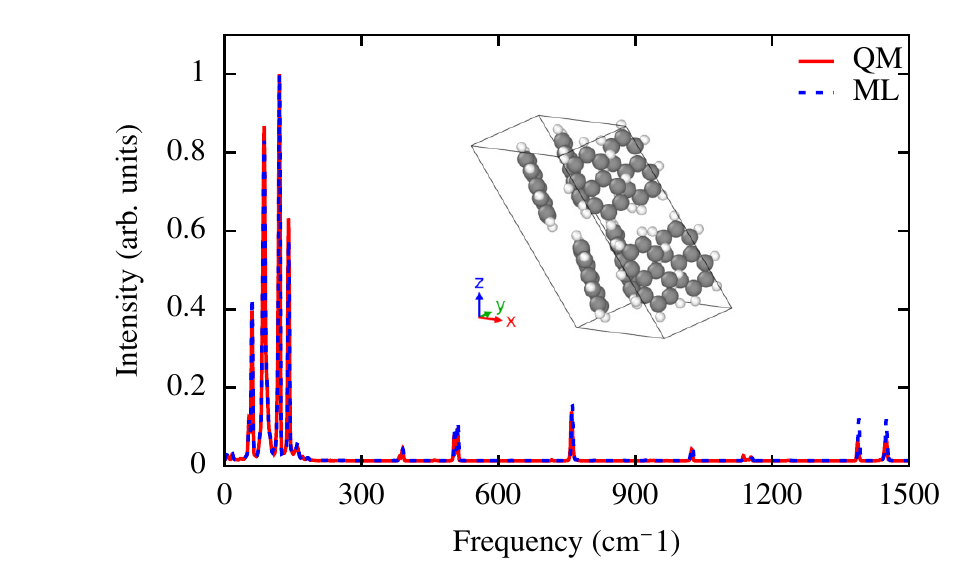}
    \caption{The powder Raman spectrum of the P2$_1$/a naphthalene crystal at 80 K, calculated  from time-correlation functions of  dielectric susceptibilities \textit{ab initio} (QM) and via an indirect SALTER prediction.}
    \label{fig:spectrum}
\end{figure}

We compared a Raman spectrum obtained using $\bm{\alpha}^{\text{QM}}$ with one obtained indirectly using SALTER predictions of $\bm{\rho}^{(1), \text{ML}}$ to obtain $\bm{\alpha}^{\text{ML}}$. The anharmonic Raman  spectrum of a disordered naphthalene sample (the powder spectrum) was obtained through ensemble-averaged time-correlation functions, as detailed in Ref.~\citenum{McQuarrie1976}. The spectra are shown in Fig.~\ref{fig:spectrum}. Good agreement is seen at the polymorph-sensitive low frequency modes, with small discrepancies in the peak intensities appearing at higher frequencies. Therefore, by applying SALTER, we obtain Raman spectrum of comparable quality to those obtained using \emph{ab initio} methods and to those obtained from a direct learning of $\bm{\alpha}^{\text{QM}}$,\cite{Raimbault2019} while dramatically reducing the simulation cost and retaining access to all the information included in the real-space distribution of $\bm{\rho}^{(1)}(\bm{r})$.

In conclusion, we presented a conceptually simple modification to the widely used $\lambda$-SOAP descriptors which allows us to use machine learning to predict a vector field; specifically, the response of the electron density to a static electric field. By adding a ``dummy'' atom which indicates the direction of the applied field to each atomic environment, we are able to retain all of the beneficial properties of the hierarchy of $\lambda$-SOAP kernels, which allowed us to predict the scalar electron density field in previous work,\cite{Lewis2021,Grisafi2022} while including (symmetry-adapted) information about the laboratory frame direction of the applied field. As a result, we expect the method developed here to be straightforward to generalise to the prediction of other vector fields. To the best of our knowledge this is the first reported machine learning model capable of predicting vector fields in a chemical context.

We applied SALTER to two condensed phase systems: liquid water and the P2$_1$/a naphthalene crystal. As expected for a disordered system, a single machine-learning model successfully predicted the density response of water to a field applied along each of the Cartesian axes; we derived the dielectric susceptibility of water from these predictions, finding values within 5\% of the reference DFPT calculations. For naphthalene, we constructed machine learning models of similar accuracy regardless of the direction of the applied field. We found a counter-intuitive relationship between the errors in the density response, which were smallest when the field was applied along the $z$ axis, and the errors in the corresponding dielectric susceptibilities, which were largest for the $yz$ and $zz$ component of the tensor. This apparent contradiction can be explained by analysing the spatial distribution of the error in the predicted density response. Nevertheless, the Raman powder spectrum obtained from these indirectly predicted dielectric susceptibilities was very similar to that obtained using DFPT, and was obtained at a fraction of the computational cost. In general, this method produces accurate derived quantities while retaining the full information contained in the electron density response to an applied field.

\section{Acknowledgements}
A.M.L. acknowledges partial support from the Alexander von Humboldt Foundation for this work. P.L and M.R. acknowledge support through the Lise Meitner Program of the Max Planck Society and the UFAST International Max Planck Research School.

\section{Data Availability}

The machine learning datasets, along with the MD trajectories and associated polarizabilities used to calculate the reference Raman powder spectrum, can be found at \url{github.com/sabia-group/SALTER-examples}. In addition, the naphthalene training data can be found in the NOMAD repository \url{dx.doi.org/10.17172/NOMAD/2023.04.13-1}. The code to build similarity kernels using the modified $\lambda$-SOAP descriptors can be found at \url{github.com/alanmlewis/TENSOAP}.

\bibliography{export}

\end{document}


\title{Supplementary Material: Predicting the Electronic Density Response of Condensed-Phase Systems to Electric Field Perturbations}
\author{Alan M.~Lewis}
\email{alan.lewis@mpsd.mpg.de}
\affiliation{Max Planck Institute for the Structure and Dynamics of Matter, Luruper Chaussee 149, 22761 Hamburg, Germany}
\author{Paolo Lazzaroni}
\affiliation{Max Planck Institute for the Structure and Dynamics of Matter, Luruper Chaussee 149, 22761 Hamburg, Germany}
\author{Mariana Rossi}
\affiliation{Max Planck Institute for the Structure and Dynamics of Matter, Luruper Chaussee 149, 22761 Hamburg, Germany}
\email{mariana.rossi@mpsd.mpg.de}

\maketitle

\section{Machine learning parameters for bulk water and naphthalene}

\begin{figure}[b]
    \centering
    \includegraphics[width=0.8\linewidth]{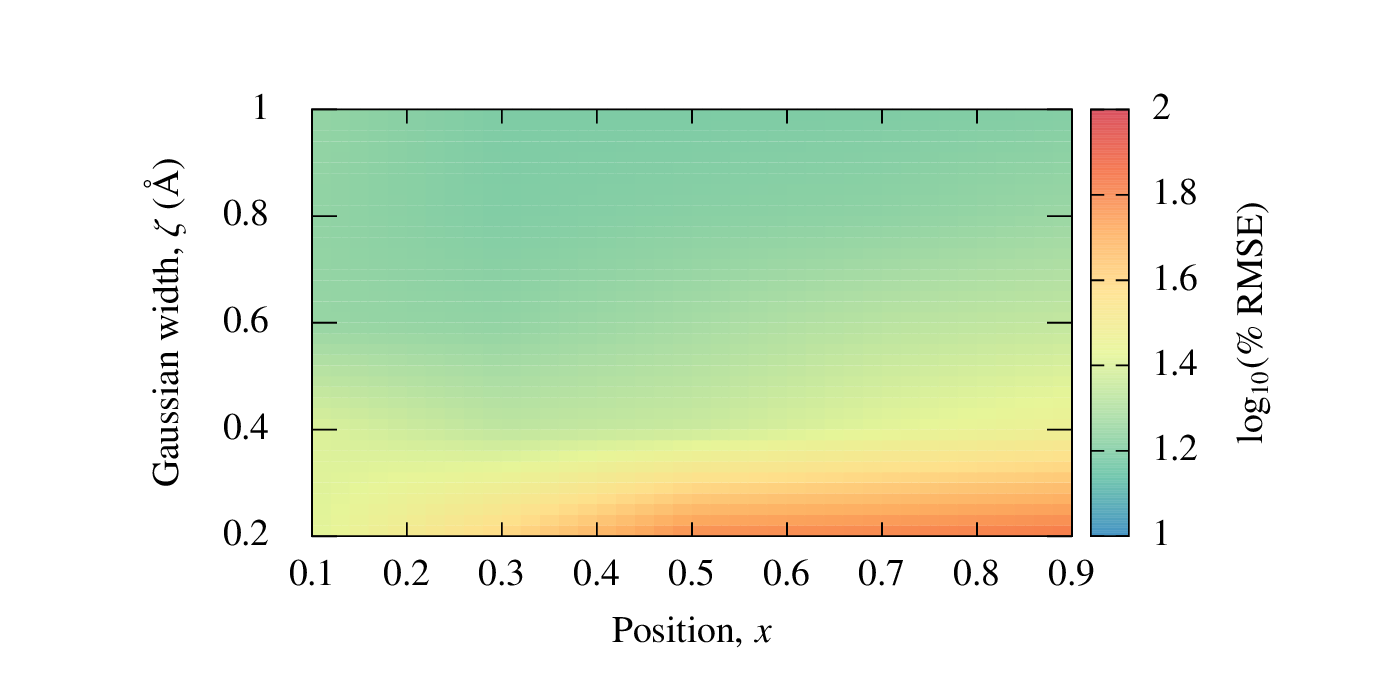}
    \caption{ The \% RMSE in the density response to a field applied along the $x$ axis for a test set of 10 water configurations, as a function of the parameters $x$ and $\zeta$ which define the position and Gaussian width associated with the dummy atom added to the atomic environment. In all cases the machine learning model is trained with $N=40$, $M=200$. }
    \label{fig:dummy_opt}
\end{figure}

In bulk water, there are physical atoms distributed throughout each spherical atomic environment $A_i$. As such, it is not clear \emph{a priori} where the dummy atom should be placed within this environment, or what the optimal width of the associated Gaussian should be. To determine these values ($x$ and $\zeta$ in Eq.~\eqref{eq:environment2} in the main text), we took a subset of 50 structures from the full training set, selected 40 of those to form a training set and used the remainder as the test set, and calculated the \% RMSE error as a function of $x$ and $\zeta$, using $M=200$. Our results are shown in Fig.~\ref{fig:dummy_opt}; we found that the error decreased monotonically with increasing $\zeta$ at all values of $x$, approaching an asymptote at large values, while for large values of $\zeta$ we found a broad minimum in $x$ centred at around 0.3. Therefore, $x=0.3$ and $\zeta = 1.0$ \AA~define the dummy atom in our calculations. Since we did not observe a deep minimum in these results, we felt confident using the same parameters for naphthalene without repeating this optimisation procedure.

\begin{figure}
    \centering
    \includegraphics[width=0.9\linewidth]{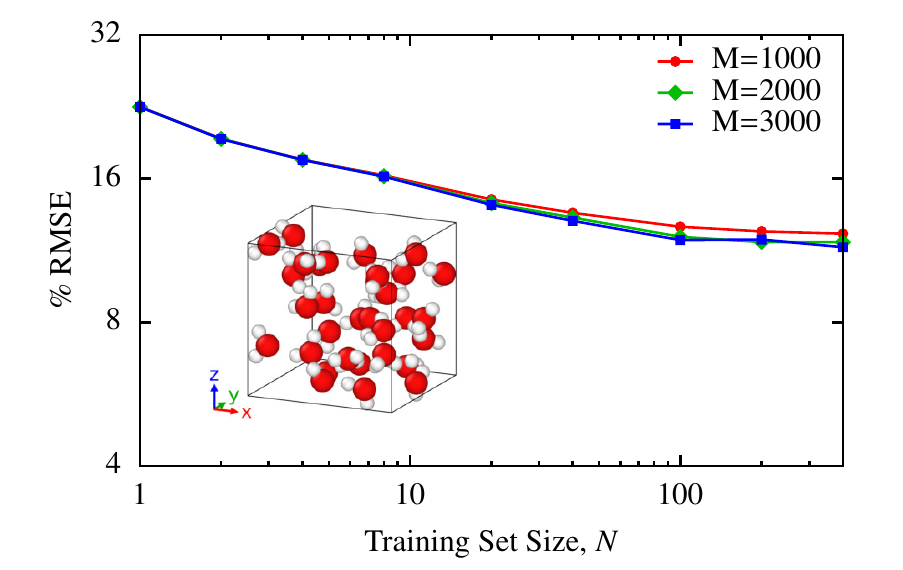}
    \caption{ Learning curves for the predicted density response to an applied field along the $x$ axis for the full bulk water dataset, converging with respect to $M$. }
    \label{fig:lcs}
\end{figure}

\begin{figure}
    \centering
    \includegraphics[width=0.9\linewidth]{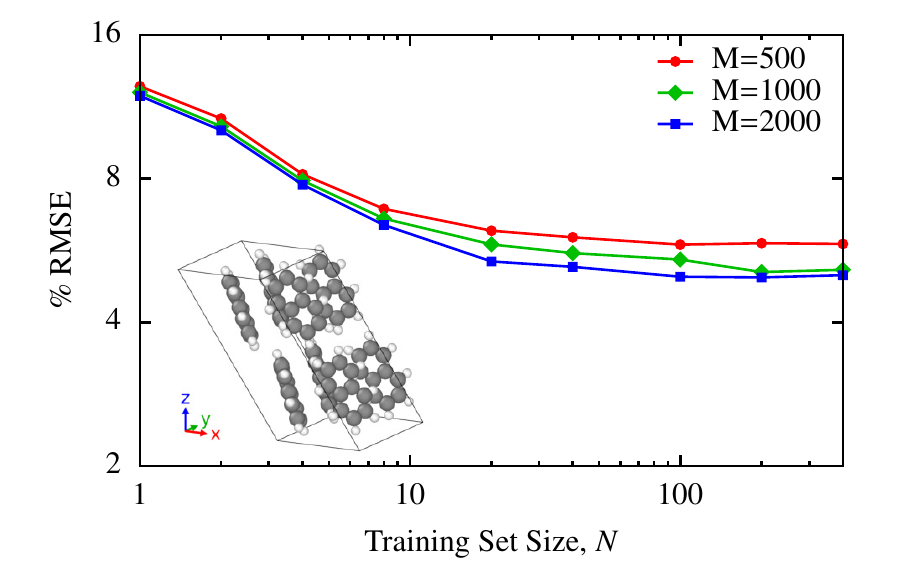}
    \caption{ Learning curves for the predicted density response to an applied field along the $x$ axis for the naphthalene dataset, converging with respect to $M$. }
    \label{fig:naph_lcs}
\end{figure}

The learning curves for the density response to a field applied along the $x$ axis are shown in Fig.~\ref{fig:lcs}, using the full dataset. As can be seen, learning improves monotonically with increasing training set size, with the error reaching a plateau for $N \ge 100$. The value of this plateau is reduced as the number of reference environments is increased, converging at $M=3000$ with errors just below 12\%. The equivalent learning curves for naphthalene are shown in Fig.~\ref{fig:naph_lcs}, showing similar behaviour, with errors converging for $M=2000$ at around 6\%.

\section{Explaining the errors in the dielectric susceptibility of naphthalene}

As noted in the main text, we observe a counter-intuitive relationship between the error in the predicted density response of naphthalene and the error in the dielectric susceptibility derived from that predicted response. Specifically, we observe the lowest errors in the predicted response to a field applied along the $z$ axis, but the largest errors in the dielectric susceptibility tensor for the components derived from this response; the reverse is true for the predicted response to a field applied along the $x$ axis. This apparent paradox can be explained in general terms by noting that the errors in properties derived from the predicted density response depend on the \emph{distribution} of the error of the density response, as well as its overall magnitude. We noted this phenomenon in our previous work predicting the electron density -- in that case we found that errors in the electrostatic energy derived from the predicted density were particularly sensitive to errors in the electron density located close to the nuclei.\cite{Grisafi2022}

\begin{figure}[t]
    \centering
    \includegraphics[width=0.8\linewidth]{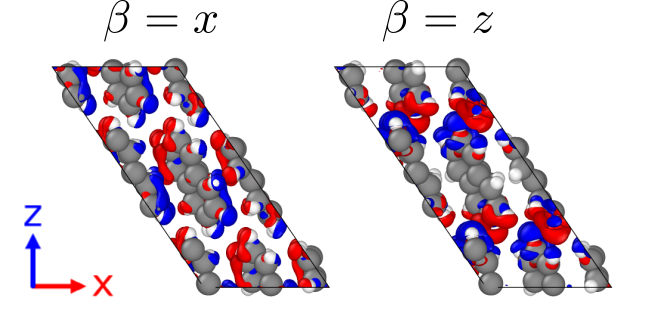}
    \caption{ Isosurfaces at $\pm$ 0.01 V\AA$^{-2}$ of the error in the predicted density response to a field applied along the $x$ axis (left) and $z$  axis (right) for a particular configuration of naphthalene. Positive errors are shown in blue, negative errors in red. }
    \label{fig:naph_err}
\end{figure}

The errors in the predicted density response to a field applied along the $x$ and $z$ axes of a representative naphthalene configuration are shown in Figure \ref{fig:naph_err}. In both cases the error in the density is localised primarily on the surfaces of the molecule normal to the applied field, with a region of positive error on one surface and of negative error on the opposing surface. Since the surface area of the molecule is largest normal to the $x$ axis, we observe a larger overall error in the density response when the field is applied along the $x$ axis than when it is applied along the $z$ axis. However, from Eq.~\eqref{eq:alpha} in the main text it can be shown that for a given magnitude of error in the density response, the magnitude of the corresponding error in the polarizability is determined by the separation between the regions of positive and negative error in the density response. To illustrate this, we imagine a simple description of an error in the density response:
\begin{equation}
\tilde{\rho}^{(1)}_\beta(\bm{r}) = \rho^{(1)}_\beta(\bm{r}) + \varepsilon\delta(\bm{r}^\prime) - \varepsilon\delta(\bm{r}^{\prime\prime}). 
\end{equation}
Here $\tilde{\rho}^{(1)}_\beta(\bm{r})$ is an approximate density response, $\rho^{(1)}_\beta(\bm{r})$ is the exact density response, $\varepsilon$ is the magnitude of the error found at two points in space $\bm{r}^\prime$ and $\bm{r}^{\prime\prime}$, and $\delta(\bm{r})$ is the Dirac delta function. There are two error terms of equal and opposite magnitude since the approximate density response must integrate to zero. The error in the polarizability arising from this error is given by
\begin{equation}
\begin{split}
        \tilde{\alpha}_{\gamma\beta} - \alpha_{\gamma\beta} = & - \int_{\textrm{u.c.}} r_{\gamma} \left(\tilde{\rho}^{(1)}_\beta(\bm{r}) - \rho^{(1)}_\beta(\bm{r}) \right) \textrm{d} \bm{r} \\
        & = \int_{\textrm{u.c.}} r_{\gamma} \left( \varepsilon\delta(\bm{r}^\prime) - \varepsilon\delta(\bm{r}^{\prime\prime}) \right) \textrm{d} \bm{r} \\
        & = \varepsilon \left(r^{\prime}_\gamma - r^{\prime\prime}_\gamma\right).
\end{split}
\end{equation}
Therefore, for a fixed value of the error $\varepsilon$, then the greater the spatial separation in the $\gamma$ direction between these regions of error, the larger the error in the polarizability and dielectric susceptibility. Since the long axis of naphthalene lies approximately along close to the $z$ axis, the separation between these regions is largest when a field is applied along the $z$ axis, and we therefore see much larger errors in the corresponding components of the dielectric susceptibility, despite the fact that the actual magnitude of the error in the density response is lowest for this field direction.

\bibliography{export}